\documentclass[lettersize,journal]{IEEEtran}
\usepackage{amsmath,amsfonts}
\usepackage{algorithmic}
\usepackage{array}
\usepackage[caption=false,font=normalsize,labelfont=sf,textfont=sf]{subfig}
\usepackage{textcomp}
\usepackage{stfloats}
\usepackage{url}
\usepackage{verbatim}
\usepackage{graphicx}
\usepackage{cite}
\usepackage{booktabs}
\usepackage{multirow}
\usepackage{amsmath}
\usepackage{mathtools}
\usepackage{amsfonts}
\usepackage{booktabs}
\usepackage{arydshln}
\usepackage{tabularx} 
\usepackage{float}

\usepackage{amsmath}
\usepackage[linesnumbered,ruled]{algorithm2e}
\usepackage{amssymb}  
\SetKwInput{KwInit}{Init}

\usepackage{booktabs} 
\usepackage{multirow} 
\usepackage{makecell}
\usepackage{subfig}

\begin{document}
\title{SDPERL: A Framework for Software Defect Prediction Using Ensemble Feature Extraction and Reinforcement Learning}  

\author{
    \IEEEauthorblockN{Mohsen Hesamolhokama$^1$\IEEEauthorrefmark{2}, Amirahmad Shafiee$^2$\IEEEauthorrefmark{1,1}, Mohammadreza Ahmaditeshnizi$^1$\IEEEauthorrefmark{1,1}, Mohammadamin Fazli$^1$, Jafar Habibi$^1$}\\
    \IEEEauthorblockA{$^1$Department of Computer Engineering, Sharif University of Technology, Tehran, Iran} \\
    \IEEEauthorblockA{$^2$Department of Mathematical Sciences, Sharif University of Technology, Tehran, Iran}\\
    {Emails:hokama@ce.sharif.edu,\{fazli, jhabibi, amirahmad.shafiee, mohammadreza.ahmaditeshnizi\}@sharif.edu}
    \thanks{*These authors contributed equally to this work.}
    \thanks{\dag Corresponding author (email: hokama@ce.sharif.edu).}
}


\maketitle

\begin{abstract}
Ensuring software quality remains a critical challenge in complex and dynamic development environments, where software defects can result in significant operational and financial risks. This paper proposes an innovative framework for software defect prediction that combines ensemble feature extraction with reinforcement learning (RL)--based feature selection. We claim that this work is among the first in recent efforts to address this challenge at the file-level granularity. The framework extracts diverse semantic and structural features from source code using five code-specific pre-trained models. Feature selection is enhanced through a custom-defined embedding space tailored to represent feature interactions, coupled with a pheromone table mechanism inspired by Ant Colony Optimization (ACO) to guide the RL agent effectively. Using the Proximal Policy Optimization (PPO) algorithm, the proposed method dynamically identifies the most predictive features for defect detection. Experimental evaluations conducted on the PROMISE dataset highlight the framework's superior performance on the F1-Score metric, achieving an average improvement of $6.25\%$ over traditional methods and baseline models across diverse datasets. This study underscores the potential for integrating ensemble learning and RL for adaptive and scalable defect prediction in modern software systems.

\end{abstract}

\maketitle

\section{Introduction}
\label{sec:sample1}

In today's rapidly evolving technological landscape, software systems serve as the backbone of critical applications across various industries, including finance, healthcare, transportation, and entertainment. Ensuring the reliability and quality of these systems is paramount, as defects can lead to costly failures, security breaches, and a loss of user trust. Despite rigorous development and testing processes, software defects are inevitable, often with significant consequences. Early prediction of these defects can substantially improve the final product's quality, reduce maintenance costs, and enhance user satisfaction \cite{fenton1999critique,catal2009systematic}.

Consider a large-scale software project involving millions of lines of code and numerous development teams. As the project progresses, the codebase evolves, new features are introduced, and existing functionalities are modified. Each change presents the potential for new defects, which can propagate throughout the system, leading to complex and unanticipated failures. Traditional defect prediction models, often based on statistical methods or conventional machine learning techniques, provide some insights by identifying defect-prone areas of the code. However, these models frequently fall short when addressing the dynamic and evolving nature of software projects, where the types and distributions of defects change over time \cite{khoshgoftaar1999logistic,arisholm2010systematic}.

A significant limitation of these traditional models is their reliance on static datasets and predefined features, which may not fully capture the complexity of evolving software environments. These models are typically trained offline, lacking the ability to continuously learn and adapt to new data. This gap indicates a need for more robust, adaptive, and accurate prediction models capable of handling the evolving nature of software development \cite{lessmann2008benchmarking,zhang2007comments}. For instance, studies have shown that incorporating more dynamic and adaptive learning approaches can enhance the accuracy of defect prediction models \cite{sutton2018reinforcement,mnih2015human}.

Reinforcement learning (RL) presents a promising solution to these challenges. Unlike traditional machine learning approaches, RL involves training an agent to make decisions by interacting with its environment to maximize a reward signal. This dynamic learning capability enables RL models to adapt to changes over time, making them particularly suited for applications with constantly changing environments, such as software development \cite{silver2016mastering,nam2017heterogeneous}. RL has demonstrated considerable success in various domains, including robotics, gaming, and autonomous systems, where adaptive decision-making is critical \cite{silver2016mastering}. The use of RL in software defect prediction (SDP), therefore, offers the potential to enhance the predictive capabilities by allowing the model to learn from ongoing development processes and adapt to emerging defect patterns \cite{ismail2024toward}.

This study proposes a novel framework for SDP that integrates ensemble feature learning with the Proximal Policy Optimization (PPO) algorithm. Ensemble feature learning enables the extraction of diverse and robust features from software data, improving the model's ability to capture complex patterns associated with software defects. The PPO algorithm, a cutting-edge method in RL, allows for continuous learning and adaptation, enabling the model to maintain high accuracy even as the software evolves \cite{schulman2017ppo}.

The contributions of this work are threefold: (1) the development of a novel SDP framework that leverages the strengths of both ensemble feature learning and RL, (2) a comprehensive evaluation of the proposed framework against traditional defect prediction approaches, demonstrating superior accuracy and adaptability, and (3) insights into the practical application of RL in the context of SDP, providing a foundation for future research.


Recent advances in artificial intelligence, particularly machine learning, have opened new avenues to improve defect prediction models. Among these, RL and ensemble feature extraction techniques have shown great promise in addressing the limitations of traditional approaches. Moreover, customization techniques, such as incorporating domain-specific configurations and utilizing auxiliary structures like pheromone tables, can further enhance prediction accuracy. Motivated by these developments, this study addresses the following research questions:

\begin{enumerate}
    \item \textbf{RQ1}: Can RL improve the performance of the SDP process?
    \item \textbf{RQ2}: To what extent do pre-trained models enhance feature extraction to boost predictive performance in SDP?
    \item \textbf{RQ3}: To what degree and in what manner can optimization of the agent's feature selection be improved?
    
\end{enumerate}

These research questions aim to investigate the potential of leveraging advanced learning techniques and novel configurations in defect prediction tasks. Addressing these questions not only provides insights into the effectiveness of these approaches but also lays the foundation for future research in SDP methodologies.

The remainder of this paper is structured as follows: Section 2 reviews related work on SDP and the application of RL in similar contexts. Section 3 details the proposed framework, including the ensemble feature learning approach and the implementation of the PPO algorithm. Section 4 presents the experimental setup and results, followed by a discussion of the findings in Section 5. Finally, Section 6 concludes the paper and outlines potential directions for future research.

\section{Related Work}
SDP has been a longstanding area of research aimed at identifying potential defects early in the software lifecycle. Traditional methods such as decision trees, support vector machines (SVM), and logistic regression were among the earliest tools applied for SDP, primarily relying on software metrics like code complexity and historical data \cite{lessmann2008benchmarking, hall2012systematic}. These methods laid the groundwork for machine learning in defect prediction but exhibited limitations in high-dimensional datasets \cite{elish2008predicting} and struggled with the challenges posed by imbalanced datasets, where defective code modules are a minority \cite{bennin2018mahakil}. Moreover, traditional methods generally focused on manually designed features, making them less effective in capturing the intricate semantic relationships within the code. Recent studies have also pointed out the limitations of these models in handling real-time, just-in-time defect prediction scenarios, which require more adaptive and robust models \cite{kamei2013}.

As traditional machine learning models faced limitations, the advent of deep learning techniques provided a significant breakthrough in the SDP domain. Unlike traditional models, deep learning models automatically extract features from raw data, making them highly effective in learning complex patterns. For instance, Wang et al. \cite{wang2018deep} developed a deep semantic feature learning model that used semantic  code features to outperform traditional feature-based models. Similarly, SL-Deep \cite{majd2020sldeep} adopted a statement-level deep learning model, using static code features to improve defect prediction. Moreover, Pornprasit and Tantithamthavorn \cite{pornprasit2023deeplinedp} explored line-level defect prediction using deep learning, further highlighting the adaptability of neural networks in defect identification. These studies emphasize that deep learning-based approaches, such as BERT \cite{uddin2022semantic}, graph neural networks \cite{xu2020defect,zhou2022graph} improve SDP performance, especially when trained on large, diverse datasets.

One critical innovation in SDP has been the development of models that leverage semantic feature learning. Traditional models, which often relied on syntactic features, struggled to capture the deeper semantic relationships in code. Recent studies have focused on incorporating semantic understanding by combining source code with external knowledge such as comments and commit messages. Liu et al. \cite{liu2023semantic} proposed a novel model that integrates both programming language data and natural language data to enrich semantic feature learning, achieving state-of-the-art performance in defect prediction. Similarly, the use of graph representation learning by Xu et al. \cite{xu2020defect} demonstrated that combining structural and semantic features via graph neural networks can improve prediction accuracy. Huo et al. \cite{huo2018learning} highlighted the importance of semantic embedding, showing how augmenting code with comments data can lead to performance improvements. These models, which blend the syntactic structure of code with its semantic context, are increasingly seen as the future of SDP, offering more reliable predictions in complex software environments \cite{ hoang2019deepjit}.

Cross-Project Defect Prediction (CPDP) is essential when there is insufficient labeled data for a target project, as it allows for knowledge transfer between projects. Several studies have explored the use of CPDP to improve the generalizability of defect prediction models. Kamei et al. \cite{kamei2016just}, for instance, conducted an extensive empirical study on just-in-time (JIT) defect prediction using cross-project models, highlighting  that JIT models learned using other projects are a viable solution for projects with limited historical data. This aligns with the work by Bai et al. \cite{bai2022transfer}, who proposed a three-stage transfer learning framework for multi-source CPDP, achieving significant performance improvements. Similarly, Majd et al. \cite{majd2020sldeep} developed the SLDeep model, which utilizes statement-level defect prediction across multiple projects, demonstrating the potential of deep learning techniques in CPDP. However, CPDP remains challenging due to the inherent differences in coding standards, project sizes, and development practices across projects\cite{aleem2023}. To mitigate these challenges, methods such as adaptive feature weighting \cite{tong2023array} and graph-based models \cite{xing2022cross} have been employed to enhance transferability. Additionally, Sheng et al. \cite{sheng2020adversarial}  proposed an adversarial discriminative convolutional neural network (ADCNN) for cross-project defect prediction, aiming to improve cross-project transferability by reducing domain discrepancies and enhancing defect prediction accuracy across different projects, while Deng et al. \cite{deng2020ast} used convolutional neural networks (CNNs) with transfer learning to tackle domain differences at the granularity of abstract syntax trees (ASTs).

One of the most persistent challenges in SDP is the class imbalance problem, where defective modules are far outnumbered by non-defective ones \cite{bejjanki2020class}. This imbalance can lead to biased models that perform well on the majority class but poorly on the minority class. Oversampling techniques such as SMOTE (Synthetic Minority Over-sampling Technique) \cite{pak2018smote} and its variants have been extensively used to address this issue. However, oversampling can lead to overfitting in some cases. To overcome this, Feng et al. \cite{feng2021coste} proposed a complexity-based oversampling technique (Coste) that focuses on more complex software modules, thereby avoiding unnecessary oversampling of simpler ones. Yedida and Menzies \cite{yedida2021value} further investigated the value of oversampling for deep learning models in SDP, concluding that class rebalancing significantly improves model performance, especially when combined with ensemble learning techniques. Zhao et al. \cite{zhao2019siamese} introduced a cost-sensitive Siamese neural network, specifically designed for defect prediction in imbalanced datasets.

Deep learning has revolutionized the field of SDP, offering models that can automatically learn complex patterns from raw code data without the need for manual feature extraction. These models, particularly CNNs and Long Short-Term Memory (LSTM) networks, have been highly successful in capturing both syntactic and semantic features from the source code \cite{wang2018deep}. For instance, the deep semantic feature learning approach by Wang et al. \cite{wang2018deep} utilized both semantic and structural information to achieve state-of-the-art performance. Similarly, deep learning methods such as SLDeep \cite{majd2020sldeep} and BiLSTM with BERT-based semantic features \cite{uddin2022semantic} have shown that deep models outperform traditional machine learning approaches, especially when handling complex high-dimensional data. Additionally, the Transformer model has shown promise in SDP. Zheng et al. \cite{zheng2021transformer} applied Transformers to capture long-range dependencies in code, significantly enhancing defect prediction accuracy. Other advanced architectures, such as GNNs (Graph Neural Networks), have been used to combine structural and semantic code features, as demonstrated by Zhou et al. \cite{zhou2022graph}, who applied GNNs to achieve superior performance in defect prediction tasks.

Hybrid models and ensemble learning approaches have proven highly effective in improving software SDP performance by combining the strengths of multiple learning algorithms. Hybrid models, which blend various deep learning architectures or combine deep learning with traditional methods, have shown significant improvements in defect detection accuracy. Tong et al. \cite{tong2018software} utilized stacked denoising autoencoders (SDAE) alongside ensemble techniques to boost defect prediction, demonstrating that combining multiple models can lead to higher robustness. Similarly, H. Tong et al. \cite{tong2024master} introduced a multi-source transfer weighted ensemble model that combined predictions from multiple models for cross-project defect prediction, significantly improving the accuracy of the ensemble model. Zhao et al. \cite{zhao2021metaheuristic} integrated metaheuristic optimization techniques with deep neural networks to improve feature selection and prediction accuracy, showing that hybrid methods can adaptively optimize the prediction process. Wu et al. \cite{wu2024novel} proposed a weighted classification model that combines association rule mining with machine learning classifiers to handle imbalanced datasets, improving defect prediction precision. In addition  Abdu et al. \cite{abdu2022deep} conducted a systematic review of deep learning-based hybrid models, emphasizing that combining semantic feature learning with hybrid models can lead to more accurate predictions. The integration of hybrid and ensemble methods provides a robust and adaptive solution to the variability and complexity of modern software systems \cite{feature_selection_sdp}.
Attention mechanisms have gained popularity in SDP due to their ability to focus on the most relevant parts of the source code, improving both interpretability and accuracy. Models that employ attention mechanisms have been able to capture long-range dependencies in code and prioritize the most defect-prone areas. Hoang et al. Yu et al. \cite{yu2021deep} applied attention mechanisms within a deep learning model, significantly improving defect prediction accuracy by emphasizing the most critical portions of the code. Yu et al. \cite{yu2021hierarchical} proposed a Defect Prediction framework based on Hierarchical Neural Networks (DP-HNN) that leverages the structure of Abstract Syntax Trees (ASTs). By segmenting file-level ASTs into subtrees centered on key nodes, DP-HNN preserves long-term dependencies and fine-grained details. This multi-granularity encoding approach improved prediction performance, outperforming state-of-the-art methods in MCC and Area Under the Curve (AUC) metrics. Zheng et al. \cite{zheng2021software} used a transformer model, incorporating attention to capture both local and global dependencies within the source code, demonstrating significant improvements over traditional deep learning models. 

In addition to these models, cost-sensitive learning has become an essential approach in SDP to minimize the cost of misclassifying defective and non-defective modules. In practical scenarios, the cost of failing to detect a defect (false negative) can be far higher than falsely flagging a clean module (false positive). Yedida and Menzies \cite{yedida2021oversampling} challenged the notion that deep learning eliminates the need for pre-processing in defect prediction. They introduced fuzzy sampling, an oversampling technique integrated into the GHOST pipeline, which optimizes hyperparameters for efficient training. Their approach demonstrated significantly improved performance over prior deep learning methods across multiple datasets, emphasizing the value of pre-processing in defect prediction.

RL has recently emerged as a promising technique in SDP, leveraging its ability to adaptively optimize the prediction process based on real-time feedback \cite{ismail2024toward}. Unlike traditional static models, RL can continuously refine its predictions as the software system evolves. Zhang et al. \cite{zhang2024deep} introduced a deep Q-learning network-based feature extraction method for defect prediction, showcasing how RL can improve the feature learning process, leading to more accurate predictions. Similarly, Ismail et al. \cite{ismail2024toward} applied deep RL to just-in-time defect prediction (JITDP), effectively reducing false positives and improving real-time prediction accuracy. Guo et al. \cite{guo2023feature} proposed a feature transfer learning (FTL) method for SDP, using RL to transfer feature knowledge between training and test data, addressing feature differences to improve prediction performance. Furthermore, Wang et al. \cite{wang2022heterogeneous} applied federated RL in heterogeneous defect prediction, using gradient clustering to enhance the adaptability of models across distributed environments. These studies highlight the potential of RL to handle complex, dynamic, and distributed software environments, making it a valuable approach for future defect prediction research.

The field of SDP has evolved significantly, from traditional machine learning methods to advanced deep learning and RL techniques \cite{kobylkin2021survey}. Early models, such as decision trees and support vector machines, laid the groundwork for defect prediction but faced limitations in handling imbalanced data and high-dimensional feature spaces \cite{thota2020survey}. The introduction of deep learning models, including convolutional neural networks and recurrent neural networks, brought automated feature extraction and improved prediction accuracy. Furthermore, semantic feature learning allowed models to capture deeper contextual relationships in the code, enhancing the precision of defect detection. whithin-project defect prediction (WPDP) and handling imbalanced datasets remain critical challenges, with methods like transfer learning and oversampling showing promise in addressing these issues. Finally, RL has emerged as a novel approach in SDP, offering dynamic, adaptive, and scalable solutions for real-time defect prediction. Together, these advancements have brought SDP closer to practical, real-world applications, ensuring that future models can better handle the complexities of modern software development.

\section{Methodology}

\subsection{Overveiw Of Method}
\begin{figure*}
  \includegraphics[width=\linewidth]{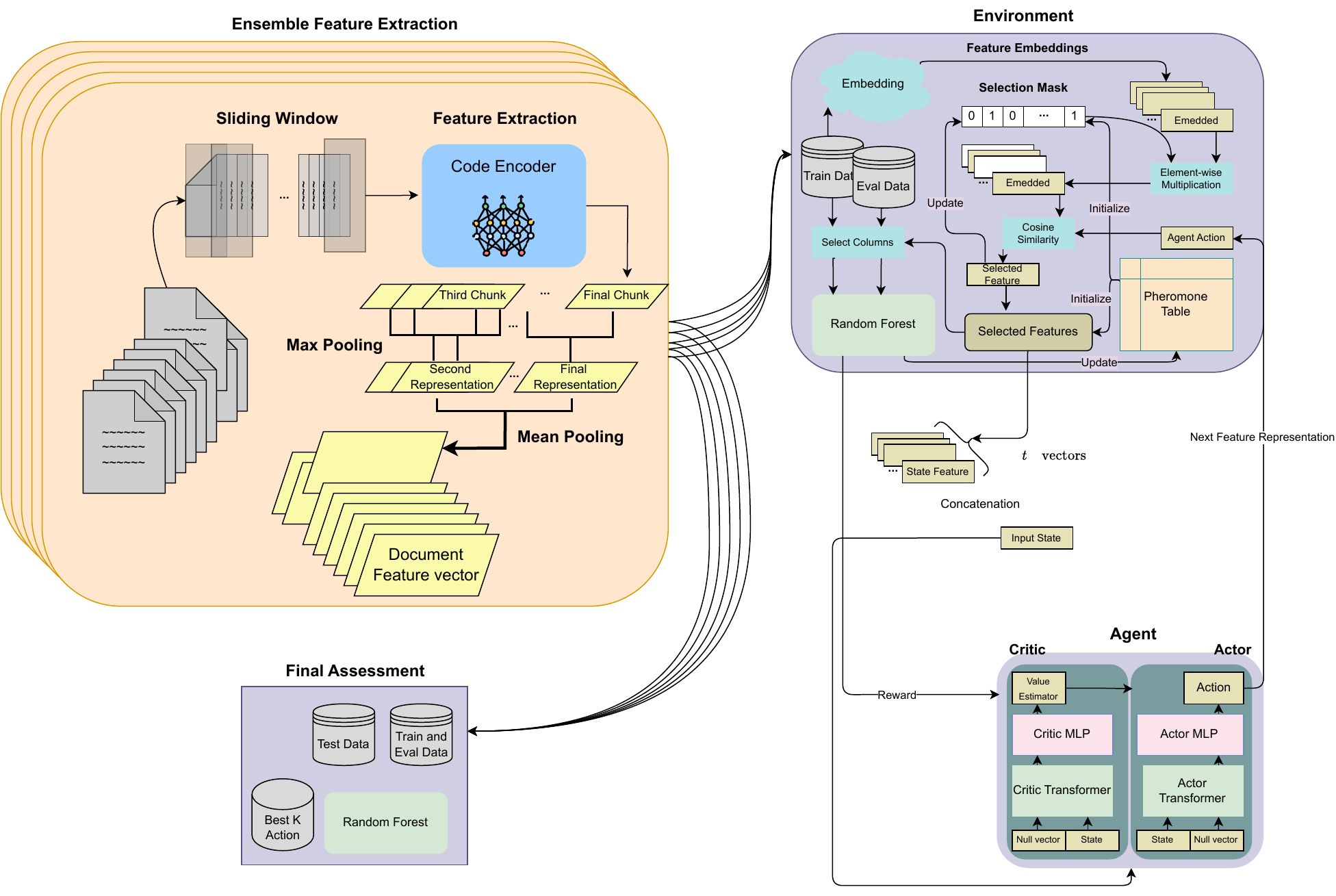}
  \caption{The Overview of SDPERL Framework}
  \label{fig:main}
\end{figure*}

In this paper, we propose an innovative framework named SDPERL, for within-project SDP, leveraging a combination of ensemble feature extraction and RL for feature selection. The proposed framework involves the use of pre-trained models such as BERT, XLNet, and CodeT5 to extract a comprehensive set of features from the source code. Each of these models captures different aspects of the code’s structure and semantics, providing a rich representation of the data. The extracted features are then subjected to an RL process using the PPO algorithm. PPO is employed to select the most significant features, thereby eliminating irrelevant or redundant features, which in turn reduces noise in the dataset and enhances the effectiveness of the classification model. Once the most effective features are selected, they are passed to a classifier, which categorizes the software files as either defective or benign. This framework, combining state-of-the-art feature extraction with dynamic feature selection through RL, aims to improve the performance of defect prediction models. An overview of the framework is shown in Figure \ref{fig:main}.

\subsection{Ensemble Feature Extraction }
The ensemble feature extraction process involves applying multiple pre-trained models to capture a wide range of features from the source code.\cite{xu2016impact} In this study, we utilize CodeBERT, GraphCodeBERT, JavaCodeBERT, XLNet, and CodeT5 to extract features from the source code.



Each of these models has been pre-trained on extensive code datasets and demonstrates an ability to comprehend both the syntactic and semantic structures of various programming languages. For example, we employed masked language models (MLMs) such as CodeBERT, which is designed to generate bidirectional representations of source code and natural language, enabling it to efficiently capture the semantic aspects of code. Similarly, GraphCodeBERT is capable of capturing the structural features of code through its graph-based representations, while JavaCodeBERT has been fine-tuned specifically for Java programs, making it more suited for tasks related to this language. In addition to MLMs, we utilized XLNet, a model based on permutation language modeling (PLM), which captures a broader range of features not addressed by traditional MLMs. Furthermore, we leveraged CodeT5’s encoder to further cover contextual characteristics of the code  through its sequence-to-sequence (seq2seq) architecture and text-to-text pretraining, thereby enriching the overall quality of the features extracted. Each of these models contributes uniquely to the analysis, offering a comprehensive understanding of both code syntax and semantics while ensuring that a wide range of features is captured for more effective and precise feature representation.

 Each of these models extracts 768 features, leading to a total of 3840 features when combined. These features represent various characteristics of the code, such as syntax patterns, semantic meanings, and code dependencies.

In addition to the features extracted by the pre-trained models, 20 static features are also collected from the code, providing a broader perspective on code characteristics. These static features include software engineering code quality metrics. These metrics can be checked out in Table \ref{tab:features}. These metrics offer valuable insights into the structural properties of the software, which are often correlated with defect proneness. By combining the semantic features extracted by the pre-trained models with the static features, we create a comprehensive feature set that captures both high-level and low-level characteristics of the software.

\begin{table*}[htbp]
\centering
\caption{Extracted Statistical Features from Code}
\resizebox{\textwidth}{!}{
\begin{tabularx}{\textwidth}{|l|X|}
\hline
\textbf{Feature} & \textbf{Description} \\ \hline
Lines of Code (LOC) & Total number of lines in the code, including blank lines and comments. \\ \hline
Source Lines of Code (SLOC) & Total number of lines that contain actual code, excluding comments and blank lines. \\ \hline
Number of Comments & 
\vspace{-5pt}
\begin{itemize}
    \item Single-line comments (`//`).
    \item Multi-line comments (`/* ... */`).
    \vspace{-7pt}
\end{itemize}\\ \hline
Comment Density & Ratio of comment lines to the total lines of code (LOC). \\ \hline
Number of Blank Lines & Total number of blank lines in the code. \\ \hline
Total Tokens & Total number of tokens (words) in the code, excluding comments. \\ \hline
Unique Tokens & Number of unique tokens in the code, excluding comments. \\ \hline
Average Line Length & Average number of characters per line of code. \\ \hline
Code Length (in characters) & Total length of the code in characters. \\ \hline
Number of Functions & Heuristic count of function or method definitions (based on common keywords like `public`, `void`, `function`, `def`). \\ \hline
Number of Variables & Heuristic count of variable declarations (based on keywords like `var`, `let`, `const`, and Java data types like `int`, `String`). \\ \hline
Number of Loops & Count of loop constructs (`for`, `while`, `do`). \\ \hline
Number of Conditionals & Count of conditional constructs (`if`, `else`, `switch`, `case`). \\ \hline
Number of Try-Catch Blocks & Count of exception-handling blocks (`try`, `catch`, `finally`). \\ \hline
Number of Imports & Count of `import` statements (Java-specific). \\ \hline
Number of Classes & Count of class declarations (`class`). \\ \hline
Number of Interfaces & Count of interface declarations (`interface`). \\ \hline
Number of Annotations & Count of annotations (`@Annotation`). \\ \hline
Number of Method Invocations & Heuristic count of method calls (e.g., `object.method()`). \\ \hline
Number of Literals & Count of string, character, and numeric literals. \\ \hline
\end{tabularx}
}
\label{tab:features}
\end{table*}


\subsection{RL for Feature Selection}
Once the features have been extracted, we employ RL to select an optimum subset of features. Given the large number of features (3840 from ensemble extraction and 20 static features), not all of them contribute equally to the prediction of software defects. Using all features can introduce noise and increase the computational complexity of the classification model. To address these, we use the PPO algorithm, a state-of-the-art RL technique known for its stability and efficiency and the ability to hold performances in high-dimensional spaces \cite{schulman2017proximal}. The main purpose of an agent in our setting would be to sequentially select an optimum subset of a given feature set, making the process of defect prediction more effective.

PPO's stable functionality over both discrete and continuous action spaces allows a variety of experimental settings, specifically the use of embedding spaces \cite{schulman2017proximal}. Feature selection is naturally a combinatorial problem modeled on a discrete space. An agent in this setting is supposed to dynamically or holistically select the optimal subset. On the other hand, having access to a representative space, each feature can be selected based on the distances from the desired representation. The agent understands the significance of each feature by engaging with an environment where its actions—selecting specific features—result in rewards determined by the performance of a classifier trained using the selected features. The reward function is designed to prioritize features that make the defect prediction process effective while penalizing the selection of redundant or irrelevant features. Over multiple episodes, the PPO agent optimizes its policy to select the most relevant features. Algorithm \ref{alg:main} describes the RL agent's training pipeline. In the following sections, we will break down the important components of this algorithm.

\subsubsection{Environment}
The environment is built upon the feature space derived from the ensemble feature extraction process. Each state in the environment corresponds to a specific subset of features extracted from the software code. The agent interacts with the environment by expanding each state to include an unselected feature. As the agent moves through the environment, it learns the significance of various features based on how effectively they contribute to enhancing classification performance in defect prediction, thus continuously adapting within the feature space.

Our environment comprises four main parts:

\textbf{Data preparation}: We load a previous version of the target project and divide it into train and validation datasets. It is also important to ensure the presence of defective data within both parts. Since the data is mostly imbalanced, we use an oversampling method (SMOTE) to bring the training dataset into equal shares of the defective and benign samples.

\textbf{Embedding}: Selecting features using their vanilla form introduces computational challenges \cite{Guyon2003}. In the simplest form, selecting an optimum subset of size $K$ from possible $N$ candidates requires $O(N^K)$ computations, due to the lack of meaningful relation between nominees \cite{Kohavi1997}. This issue, however, does not hold in situations when we deal with embedding spaces or meaningful relations \cite{Bengio2013}, for an optimal solution generally consists of eigenvectors of the goal subspace \cite{belkin2003laplacian}.

We tackle this challenge by creating a representative space for features, establishing semantics in the feature space, and reducing the computational complexity. We achieve this by utilizing the unsupervised method, K-means, and the distribution of each feature within the training data. Initially, we assign each feature $f_i$ to a statistical four-dimensional vector $s_i$ that represents the means and variances for both defective and benign data. Next, we cluster these statistical vectors into $k$ groups for various values of $k$. We hypothesize that for every value of $k$ in the K-means algorithm, the separation is done based on different hidden characteristics of data. We hence run the K-Means algorithm and categorize features for different values of $k$, obtaining a one-hot vector for each feature in each iteration. Next, for each feature, we concatenate each of the obtained one-hot encodings with the initial statistical vector, resulting in its final embedding. Formally, an embedding of feature $f_i \in F$ is represented as $v_i$ where:
\vspace{-5pt}
\begin{align}
    &v_i = U(s_i) \mathbin\Vert s_i \\
    &s_i = [\mathrm{E}_0[f_i], \mathrm{Var}_0[f_i], \mathrm{E}_1[f_i], \mathrm{Var}_1[f_i]]\\
    &U(s_i) = [KM(s_i)_{k_1}\mathbin\Vert \cdots\mathbin\Vert KM(s_i)_{k_K}]
\end{align}

\begin{figure}
  \includegraphics[width=\linewidth]{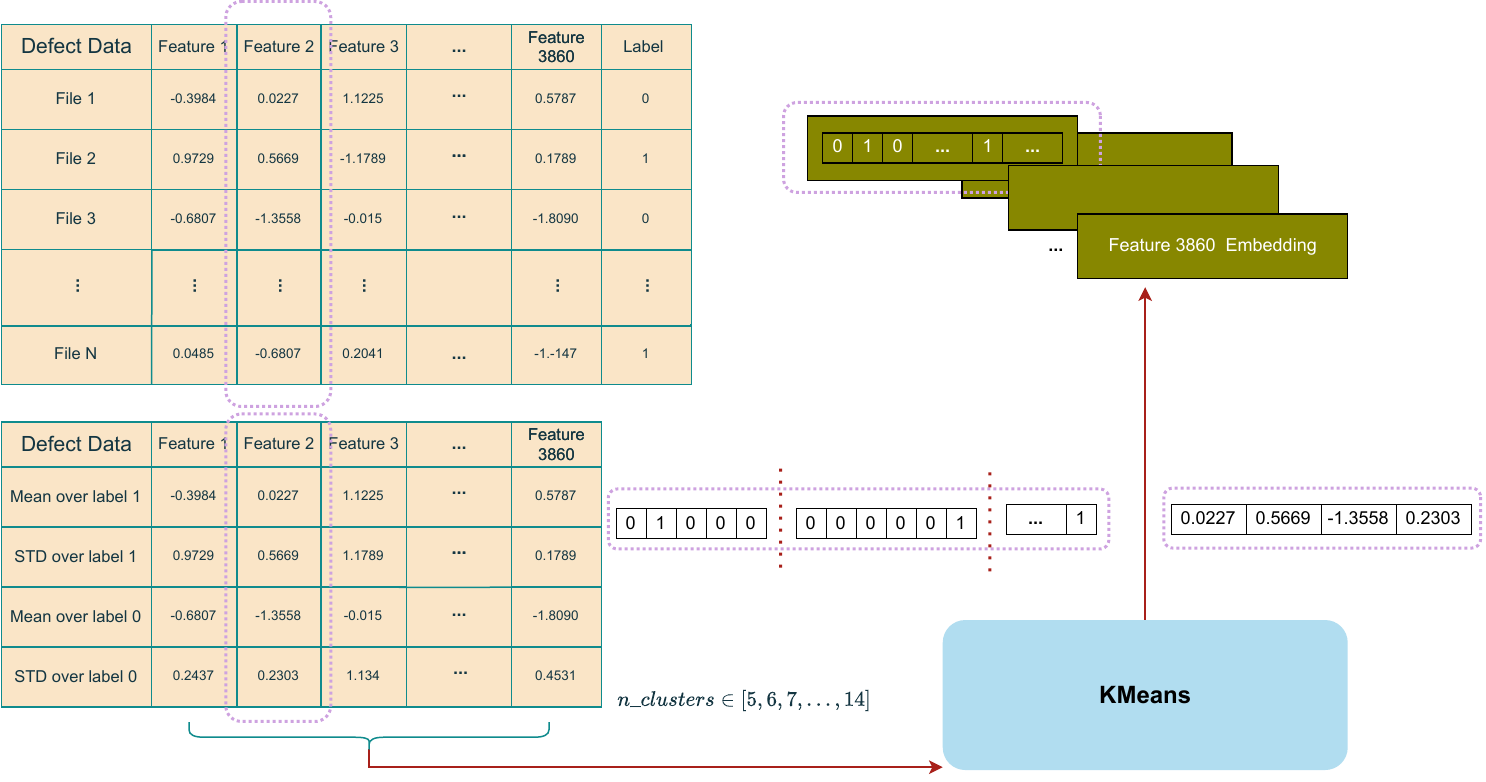}
  \caption{Embedding Pipeline}
  \label{fig:embedding}
\end{figure}

In which $KM(.)_{k_j}: S \rightarrow \mathbb{R}$ is a function with the domain of all statistical vectors. For a given statistical vector $s_i \in S$, this function returns the one-hot vector of the corresponding category of $s_i$ as defined by the K-means clustering with \( k = k_j \). The terms \( \mathrm{E}_0\), \(\mathrm{Var}_0\), \(\mathrm{E}_1 \), and \( \mathrm{Var}_1 \) represent the means and variances of the feature across labels benign and defective data, respectively. This process can be observed in Figure \ref{fig:embedding}.

\textbf{Training Black Box}: After each step, when a new action is selected by the agent, it is essential to examine its quality. We use a classifier model for this task, training the model on the training data and testing it on the evaluation data with only the selected features.

\begin{figure}
  \includegraphics[width=\linewidth]{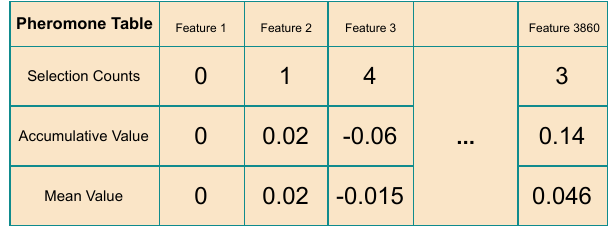}
  \caption{Structure Of Phoromone Table}
  \label{fig:pheromone}
\end{figure}

\textbf{Pheromone Table}: Inspired by ACO \cite{dorigo2006ant}, we use a pheromone table to store the quality of each feature in SDP. In each iteration, for the selected feature $f_i$, we update the corresponding entry in the pheromone table with the TD reward. formally the pheromone level of feature $f_i$ after timestep $t$ is calculated as $PH^t_i = ph^t_i / c^t_i$, where $PH^t_i$ represents the \( i \)-th element of the pheromone table \( PH \) after timestep \( t \). Here, \( ph \) denotes the vector of accumulated TD rewards, and \( c_i \) corresponds to the number of times feature \( f_i \) has been selected. An illustration of the pheromone table is shown in Figure \ref{fig:pheromone}.

\subsubsection{Action}
In a classical feature selection problem, an agent takes an action as a direct selection from the available feature set. Provided by the specific embedding step taken in the environment, the agent is equipped with a solid understanding of the feature set's underlying semantical relation. Therefore, feature selection in each step is reduced to defining the embedding of a preferred action.

The RL agent deals with the representations introduced in the environment section. At each time step, the already selected features by the model are passed as a current state in the format of the representative vectors i.e. for timestep $t$, the input state is a set of size $t-1$ vectors. As an action, the model is supposed to give back a desired representative vector. We hypothesize that during the iterations, the agent can be capable of understanding the relations between each of the clustering settings and groups. Hence, it should be able to determine the best representation to complement an already selected set. As a result of this hypothesis, we measure the Euclidean distance of the agent's action to the non-selected features' vector representations and choose the closest options. Since we aim to provide a clearer reward signal to the model, we avoid any randomness in selecting the closest feature.

\subsubsection{Reward}
After each action, the selected features are fed into the classification model, and the prediction performance determines the reward. Higher rewards are given for feature sets that lead to better classification performance on the validation dataset. This reward signal leads the model towards high performance without pronation to overfitting.

As mentioned in the environment section, we utilize a classifier model as our quality examiner or in other words, the reward function. After the arrival of each new action by the RL agent, we add this new feature to the already selected set and separate the corresponding portion of the training and evaluation set. We then pass these data to our classifier and receive the results. We do not apply any hyperparameter optimization here as the classifier model is to be the black box reward function. The quality of the new action is examined via TD from the outcoming criteria received by the classifier's reported performance over the evaluation data.

Upon receiving a reward \( r \), PPO updates its policy by calculating the advantage function \( A_t \), which reflects the difference between the observed reward \( r \) and the expected value of the current state. This advantage guides the update process as follows:

\textbf{Advantage Calculation:} The reward \( r \) is used to estimate the advantage \( A_t = r + \gamma V(s') - V(s) \), where \( V(s) \) is the state-value function. This quantifies how much better the taken action was compared to the expected outcome in the current policy \cite{schulman2017ppo, schulman2018highdimensionalcontinuouscontrolusing}.

\textbf{Policy Adjustment:} Using \( A_t \), the agent computes the probability ratio \( r_t(\theta) = \frac{\pi_\theta(a_t|s_t)}{\pi_{\theta_\text{old}}(a_t|s_t)} \). The policy is updated by maximizing the clipped surrogate objective presented in equation \ref{eq:ppo},    where the reward \( r \) influences the size and direction of policy changes \cite{schulman2017ppo, schulman2018highdimensionalcontinuouscontrolusing}.
\vspace{-5pt}
\begin{equation}\label{eq:ppo}
        L(\theta) = \mathbb{E}[\min(r_t(\theta)A_t, \text{clip}(r_t(\theta), 1-\epsilon, 1+\epsilon)A_t)]
\end{equation}

\textbf{Stability through Clipping:} The clipping mechanism ensures the policy update remains constrained, preventing overly aggressive changes even for large \( r \)-induced advantages. This stabilizes learning while still leveraging \( r \) to improve policy performance \cite{schulman2018highdimensionalcontinuouscontrolusing}. Compatible with the problem, this property can prohibit drastic changes in the policy regarding specific state-action pairs which can be crucial partly due to the correlations among different features.

\subsubsection{Decision Network}

The decision network in PPO plays a critical role in defining the agent’s overall policy during feature selection. It is typically composed of a policy network and a value network. The policy network determines the probability distribution over possible actions based on the current state of the environment, while the value network estimates the expected reward for the current state. These networks help the agent to estimate the value of each feature representation, allowing it to make informed decisions on feature selection as it seeks to maximize the cumulative reward over time.

Our agent consists of two separate networks of actor and critic, where the actor network is responsible for action production and the critic is the value network, responsible for state valuation. Based on the nature of our work, we have built our networks on the structure of transformers \cite{vaswani2017attention}. In each network, the input state to the transformer module is augmented with an empty tensor. The corresponding output of the transformer module to these tensors will be the representative of the desired feature, in the actor-network, and the current state's value, in the critic network. We mask the attention layers of each timestep so that the attention from the empty tensor to other state elements is blocked. Both actor and critic networks employ and MLP structure over their corresponding empty tensors output from the transformer to reach the desired feature representation and state value estimation, respectively.

\begin{algorithm}[]
\caption{RL Agent Training Loop}
\label{alg:rl_classifier}
\KwIn{Data $\mathcal{D}$, Timesteps $\mathcal{T}$, Max\_features $\mathcal{M}$, Num\_features $N$, Embedding\_low $k_\text{start}$, Embedding\_high $k_\text{end}$}
\KwInit{Randomly split $\mathcal{D}$ into training set $\mathcal{D}_{train}$ and evaluation set $\mathcal{D}_{eval}$ with seed $s$.}

\While{$\mathcal{D}_{eval}$ does not contain label 1}{
    Randomly resplit data with incrementing seed $s$\;
}

Compute embedding: $\mathcal{E}, embed\_dim = \texttt{Embedder}(\mathcal{D}, k_\text{start}, k_\text{end})$\;
Initialize pheromone table $\mathcal{PH} = [(0,0), \dots, (0,0)]$ ($\mathcal{N}$ pairs)\;
Initialize RL agent\;
Initialize Classifier\;
Select $\lfloor\mathcal{M}/3\rfloor$ features from pheromone table $\mathcal{PH}$: $\mathcal{V}$\;

Initialize state $\mathbf{s}_0 = \mathcal{V} + [\mathbf{0}_{embed\_dim}, \dots, \mathbf{0}_{embed\_dim}]$ (totaling $\mathcal{M}$ features)\;

\textbf{Main Training Loop:}\\
Call \texttt{Train}($\mathcal{D}_{train}$, $\mathcal{D}_{eval}$, $\mathcal{E}$, $embed\_dim$, $\mathcal{T}$, $\mathcal{M}$) to perform RL-based training and update agent;

\textbf{Post-training:}\\
Load best action set $A_{\text{best}}$\;
Train classifier on entire dataset $\mathcal{D}$ using $A_{\text{best}}$\;
Test classifier on test data $\mathcal{D}_{test}$\;
\end{algorithm}

\begin{algorithm}[]
\caption{Train}
\label{alg:train}
\KwIn{Training set $\mathcal{D}_{train}$, Evaluation set $\mathcal{D}_{eval}$, Embedding $\mathcal{E}$, Embedding Dimension $embed\_dim$, Timesteps $\mathcal{T}$, Max\_features $\mathcal{M}$}
\KwOut{Best Action Set $A_{\text{best}}$}

Initialize Action set $\mathcal{A} = \emptyset$, Reward $\mathcal{R} = 0$, Cumulative Reward $\mathcal{CR} =0$\;

\For{each timestep $t = 1$ to $\mathcal{T}$}{
    Agent receives state $\mathbf{s}_t$\;
    Agent takes action $\mathbf{a}_t$\;
    
    Environment calculates distance between $\mathbf{a}_t$ and rows in $\mathcal{E}$\;
    Environment selects $\mathbf{a}'_t = \arg\min\limits_{\mathbf{e} \in row_\mathcal{E}} ||\mathbf{a}_t - \mathbf{e}||$\;
    Identify feature $\mathbf{f}_t$ corresponding to $\mathbf{a}'_t$\;
    
    Append $\mathbf{f}_t$ to action set $\mathcal{A}$\;
    Train classifier on $\mathcal{D}_{train}$ using action set $\mathcal{A}$\;
    Evaluate classifier on $\mathcal{D}_{eval}$\;
    
    Classifier provides reward $\mathcal{CR}_t = \mathbf{r}_t$\;
    Compute TD reward $\mathcal{R}_t = \mathcal{CR}_t - \mathcal{CR}_{t-1}$\;
    
    Update agent with $\mathcal{R}_t$\;
    Update pheromone table $\mathcal{PH}$ on $\mathbf{f}_t$: $\mathcal{PH}(\mathbf{f}_t) += (\mathcal{R}_t, 1)$\;
    
    \If{$t \mod \mathcal{M} = 0$ (end of episode)}{
        Reinitialize state $\mathbf{s}_0$, action set $\mathcal{A} = \emptyset$, rewards $\mathcal{R} = 0$\;
        \If{$\mathcal{A}$ is the best action set so far}{
            Save best action set $A_{\text{best}}$\;
        }
    }
}
Return $A_{\text{best}}$\;
\end{algorithm}\label{alg:main}

\begin{algorithm}[]
\caption{Embedder}
\label{alg:clustering_one_hot}
\KwIn{Data $\mathcal{D}$, Range of clusters $k_{\text{start}}$ to $k_{\text{end}}$}
\KwOut{Clustered Data $\mathcal{E}$, Embedding size $|\mathcal{E}|$}

\textbf{Preprocessing} \\
Remove the \texttt{'Bug'} column from the dataset $\mathcal{D}$\;
\textbf{Filter} $\mathcal{D}$ into two sets: $\mathcal{D}_{label_1}$ (where 'Bug' = 1) and $\mathcal{D}_{label_0}$ (where 'Bug' = 0)\;

\textbf{Compute Statistics} \\
\ForEach{label $l \in \{0, 1\}$}{
    Calculate the mean of $\mathcal{D}_{label_l}$\;
    Calculate the variance of $\mathcal{D}_{label_l}$\;
}
\textbf{Store} the results in a DataFrame $\mathcal{S}$ for later use\;

\textbf{Initialize Data Structures} \\
Define cluster values $k$ in the range [$k_{\text{start}}$, $k_{\text{end}}$]\;
Initialize an empty DataFrame $\mathcal{O}_{\text{one\_hot}}$ for one-hot encoded vectors\;
Initialize an empty DataFrame $\mathcal{L}_{\text{cluster\_labels}}$ to store cluster labels for each $k$\;

\textbf{Clustering and One-Hot Encoding} \\
\ForEach{$k \in [k_{\text{start}}, k_{\text{end}}]$}{
    Apply K-Means clustering with $k$ clusters to $\mathcal{S}^\top$\;
    \textbf{Store} the resulting cluster labels in $\mathcal{L}_{\text{cluster\_labels}}[k]$\;
    
    \textbf{Generate} one-hot encoded vectors for the cluster labels\;
    \textbf{Store} the one-hot vectors in $\mathcal{O}_{\text{one\_hot}}$\;
}

\textbf{Combine Results} \\
Combine the one-hot encoded vectors $\mathcal{O}_{\text{one\_hot}}$ with the statistical data $\mathcal{S}$\;
\textbf{Final Output} \\
Set $\mathcal{E} = \mathcal{O}_{\text{one\_hot}} + \mathcal{S}$\;
$|\mathcal{E}| = \text{number of columns in } \mathcal{E}$\;

\end{algorithm}



\section{Results}
In this section, we will provide our observations and results from our experiments. We will break our observations into three settings: efficiency of unsupervised embedding, pheromone table, and their combination. We will then discuss our observations over different datasets and bold specific findings.

\subsection{Experimental Setup}
We used Kaggle’s CPU-only environment for our experimental setup, which provides 16 GB of RAM and around 20 GB of persistent storage. The CPU is an Intel Xeon with 2 cores and a clock speed of up to 2.3 GHz. Each experiment was run for 30,000 timesteps, empirically providing robust results while maintaining cost efficiency.
For our agent, we used the default SB3\textsuperscript(1) parameters. For our decision network, over all of the experiments we maintain the introduced transformer architecture, consisting of four stacking layers each, with four attention heads and the hidden dim of 16.

\subsection{Dataset} 
In this study, we used the PROMISE dataset to evaluate SDPERL for SDP. PROMISE is a widely used repository that provides defect-related data from various real-world software projects. This ensures a reliable and comprehensive benchmark for evaluating defect prediction models.

The specific projects selected for our experiments include:
    \textit{Groovy},
    \textit{Hive},
    \textit{Derby},
    \textit{Camel},
    \textit{HBase},
    \textit{ActiveMQ},
    \textit{Wicket}
     and \textit{Lucene}.

The defect data for these projects was retrieved from the LINE Dataset repository, available on GitHub \footnote{https://github.com/awsm-research/line-level-defect-prediction}. The inclusion of these projects from PROMISE ensures the diversity of our experiments across different software domains and development environments.

By leveraging the PROMISE dataset, our study benefits from high-quality, publicly available datasets that facilitate reproducibility and comparative analysis in defect prediction research.

\subsection{Performance Metrics}

Evaluating the performance of SDP models requires robust metrics that can capture different aspects of prediction quality, especially given the imbalanced nature of many defect prediction datasets. Commonly used metrics include precision, recall F1-score, and AUC. Each metric provides unique insights into the model’s effectiveness, especially in terms of balancing the trade-off between false positives and false negatives, which is critical in defect prediction contexts.

\subsubsection{F1-Score}


The F1-score, the harmonic mean of precision and recall, is crucial in defect prediction for handling class imbalance. It balances the model’s ability to correctly identify defective files (precision) and detect all actual defective files (recall), providing a robust measure for high-stakes software engineering scenarios where rare but critical defects matter. F1-score is often prioritized in defect prediction as it better reflects the model's robustness in identifying rare but critical defective instances \cite{tantithamthavorn2016automated, arar2018comprehensive}.

\subsubsection{AUC} 

AUC is a key metric in defect prediction, assessing a model's ability to distinguish between defective and benign classes across thresholds. It balances recall and false positive rates, with scores closer to 1 indicating stronger performance. AUC is especially useful in imbalanced datasets, offering a comprehensive evaluation of classifier performance under varying conditions. \cite{xu2016mining, tantithamthavorn2015impact}.



\subsection{Experimental Settings}


This section offers a comprehensive overview of our experimental configuration aimed at evaluating the effectiveness of our proposed method. We begin with an examination of the embedding space experiments under Simple and Custom configurations. Subsequently, we examine pheromone table setups, such as Vanilla, Pheromone, and Pheromone-Best-Action. 


\subsubsection{Simple Setting}
In this setting the model is aimed to choose sequentially from all possible features. For this purpose, we represent features via one-hot encoding in the state space. Furthermore, the action space is defined by a $3860$ sized vector representing the scores given by the agent to each of the features for further addition to the presented set. The underlying mechanisms of the agent remain the same in all settings.

At each timestep $t \in [1,20]$, the agent is provided with $t-1$ one-hot vectors as the already selected set of features. The next feature will then be selected from the scores provided by the agent, deterministically, from the available features. Details regarding the reward signals and decision network will be given in the following sections.

\subsubsection{Custom Embedding}
In contrast to the simple setting, in this experiment, the agent is not directly involved with the feature space. As discussed earlier, We provide the agent with a representation space of the feature space. For this aim, we iteratively cluster features into $5$ up to $14$ classes. Our empirical results show these endpoints to be the most effective. As a result, each feature will be represented by a vector of size 99, consisting of one-hot vectors from the clustering and a four-sized vector of means and variances over labels defective and benign.

Agent behavior is similar to the Simple setting. At each time-step $t \in [1,20]$, the agent is provided with $t-1$ feature representation vectors and aims to take action in the format of a vector of size $99$, representing its desired feature. We then calculate the Euclidean distances of feature representations to this action and select deterministically the one with the shortest distance.

\subsubsection{Pheromone}
As described in the methodology section, in each timestep and after each action selection, the pheromone table is updated. Supposing that $f_t$ is the selected feature at time-step $t$, we will update the corresponding column of this feature by incrementing its first row by a unit and its second row by the TD reward received from the classifier model.

To use this table more effectively and to help the agent with its action history, for each episode we initiate the state space with a set of actions sampled from this table stochastically and based on its third row i.e. features with more evidence of positive effect. We initiate a third of each episode length each time. The use of a pheromone table is not contingent on the other modules' functionality, therefore, it can be used with both Simple and Custom settings.

Naturally, using the pheromone table, the top 20 features with the highest pheromone levels are expected to form the best set of features. Therefore, as the most straightforward approach, we take the top 20 best features, accordingly, during the test phase. We will recall this approach as Pheromone in our experiments.
\subsubsection{Pheromone-Best-Action}
This setting shares all similarities to the Pheromone setup except the behavior during the test phase. In this phase, instead of using the top 20 features from the table directly, We refer to the history of computations of our agent and select the best action based on the evaluation phase performance.

\subsubsection{Vanilla}
In contrast to the Pheromone and Pheromone-Best-Action approaches, the Vanilla setting omits the pheromone table, implementing each Simple or Custom configuration independently of this technique.

\subsection{Comparison with Baseline Approach}
To evaluate the effectiveness of our proposed method, we compare it against a prominent baseline approach for SDP. The baseline method, described by \cite{wang2018deep}, leverages deep semantic feature learning using a Deep Belief Network (DBN) to extract semantic representations from code. This method addresses limitations in traditional feature extraction by encoding both the syntactic and semantic information of code segments. Specifically, the baseline model uses abstract syntax trees to create token representations of code, which are then processed through DBN layers to generate high-level semantic features. These features serve as inputs to defect prediction models, which have been shown to perform well in both file-level and change-level prediction tasks within and across projects \cite{wang2018deep}.

By contrasting our ensemble-based, RL-enhanced feature selection approach with the DBN-based baseline, we aim to highlight improvements in model performance and the ability to handle diverse codebases. The following section presents a table that compares our method and the baseline method in terms of feature extraction techniques, feature selection processes, and overall predictive capability.

\subsection{Performance Evaluation}

\begin{figure}
  \includegraphics[width=\linewidth]{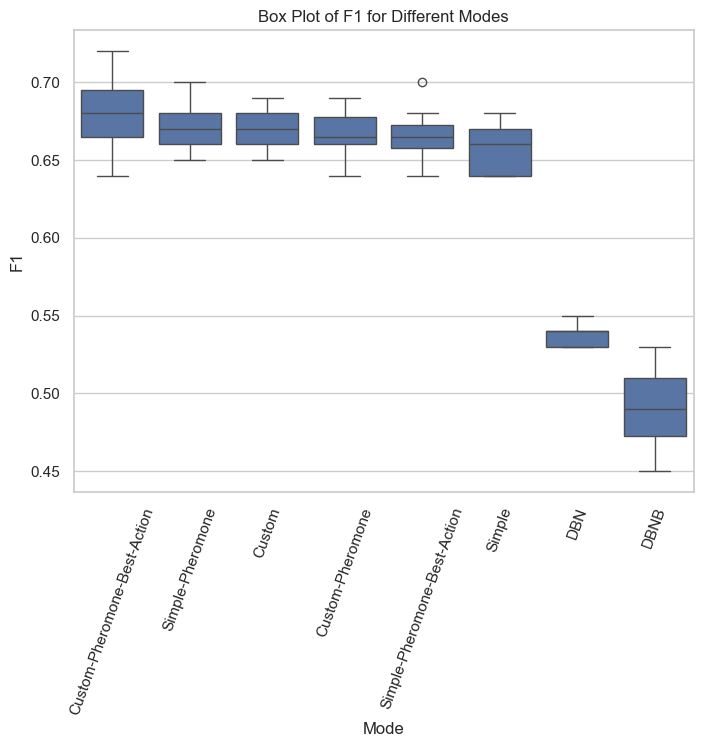}
  \caption{F1-scores In Different Modes}
  \label{fig:baseline}
\end{figure}

Figure \ref{fig:baseline} demonstrates that our proposed methods consistently outperform the baseline approaches in the Lucene dataset, with Custom-Pheromone-Best-Action emerging as the top performer in terms of F1-Score. We also evaluated this performance across various datasets found in the literature. Table \ref{tab:comparison_all_datasets_compact} further supports these findings. Custom-Pheromone-Best-Action maintains its superiority across a wide range of datasets, with only \textit{Wicket} being the exception. Additionally, we observe a strong performance across the AUC and Precision metrics, while achieving comparable results in Recall. These findings highlight the robustness and balance of our proposed method.

\begin{table*}[htbp]
\centering
\renewcommand{\arraystretch}{1.3} 
\setlength{\tabcolsep}{4.5pt} 
\begin{tabular}{|c|c|cccc|cccc|cccc|cccc|}
\toprule
\multirow{2}{*}{\textbf{Data}} & 
\textbf{Version} & 
\multicolumn{4}{c|}{\textbf{ALL}} & 
\multicolumn{4}{c|}{\textbf{Custom-Ph-Best-Action}} & 
\multicolumn{4}{c|}{\textbf{DBN}} & 
\multicolumn{4}{c|}{\textbf{DBNB}} \\ 
\cmidrule{2-18}
&Train $\rightarrow$ Test & Prec & Rec & F1 & AUC & Prec & Rec & F1 & AUC & Prec & Rec & F1 & AUC & Prec & Rec & F1 & AUC \\ 
\midrule
Groovy& $1.5.7 \rightarrow 1.6.2\beta$  & \textbf{0.73} & 0.67 & \textbf{0.73} & \textbf{0.859} & \underline{0.72} & \textbf{0.68} & \underline{0.70} & \underline{0.809} & 0.70 & 0.54 & 0.57 & 0.545 & 0.68 & \textbf{0.68} & 0.68 & 0.676 \\ 
Hive   & $0.10.0 \rightarrow 0.12.0$ & \underline{0.66} & 0.56 & \underline{0.58} & \textbf{0.790} & 0.64 & \underline{0.59} & \textbf{0.61} & \underline{0.773} & \textbf{0.71} & 0.52 & 0.51 & 0.517 & 0.56 & \textbf{0.66} & 0.55 & 0.661 \\ 
Derby & $10.3.1.4 \rightarrow 10.5.1.1$  & \textbf{0.64} & \textbf{0.81} & \textbf{0.68} & \textbf{0.879} & \underline{0.62} & \underline{0.79} & \underline{0.65} & \underline{0.872} & 0.59 & 0.70 & 0.61 & 0.701 & 0.57 & 0.74 & 0.56 & 0.736 \\ 
Camel  & $2.10.0 \rightarrow 2.11.0$ & \textbf{0.57} & 0.55 & 0.56 & \underline{0.798} & \textbf{0.57} & \underline{0.58} & \textbf{0.58} & \textbf{0.803} & \underline{0.56} & \underline{0.58} & \underline{0.57} & 0.583 & 0.52 & \textbf{0.64} & 0.52 & 0.636 \\ 
Hbase &  $0.95.0 \rightarrow 0.95.2$ & \underline{0.60} & 0.61 & \textbf{0.60} & \textbf{0.723} & 0.59 & \textbf{0.62} & \textbf{0.60} & \underline{0.702} & \textbf{0.63} & 0.53 & 0.54 & 0.533 & 0.54 & \textbf{0.62} & 0.51 & 0.623 \\ 
Activemq& $5.3.0 \rightarrow 5.8.0$  & \textbf{0.66} & {0.68} & \textbf{0.67} & \textbf{0.884} & 0.62 & \textbf{0.70} & \underline{0.65} & \underline{0.879} & \underline{0.64} & 0.62 & 0.63 & 0.617 & 0.57 & \underline{0.69} & 0.59 & 0.693 \\ 
Wicket& $1.3.0 \rightarrow 1.5.3$  & \underline{0.54} & \underline{0.69} & \underline{0.54} & \textbf{0.819} & {0.53} & 0.67 & 0.51 & \underline{0.813} & \textbf{0.57} & 0.53 & \textbf{0.55} & 0.535 & 0.52 & \textbf{0.71} & 0.41 & 0.708 \\ 
Lucene& $2.9.0 \rightarrow 3.0.0$ & \textbf{0.68} & \underline{0.75} & \underline{0.71} & \textbf{0.900} & \textbf{0.68} & \textbf{0.77} & \textbf{0.72} & \underline{0.884} & 0.54 & 0.54 & 0.54 & 0.539 & 0.52 & 0.57 & 0.52 & 0.573 \\ 
\bottomrule
\end{tabular}
\caption{Comparison of Custom-Pheromone-Best-Action, DBN, DBNB, and ALL Across All Datasets (The best result for each metric-dataset pair is shown in \textbf{Bold}, and the second-best result is \underline{underlined}.)}
\label{tab:comparison_all_datasets_compact}
\end{table*}

\section{Discussion}

\subsection{What is the most effective feature set size?}
\begin{figure}
  \includegraphics[width=\linewidth]{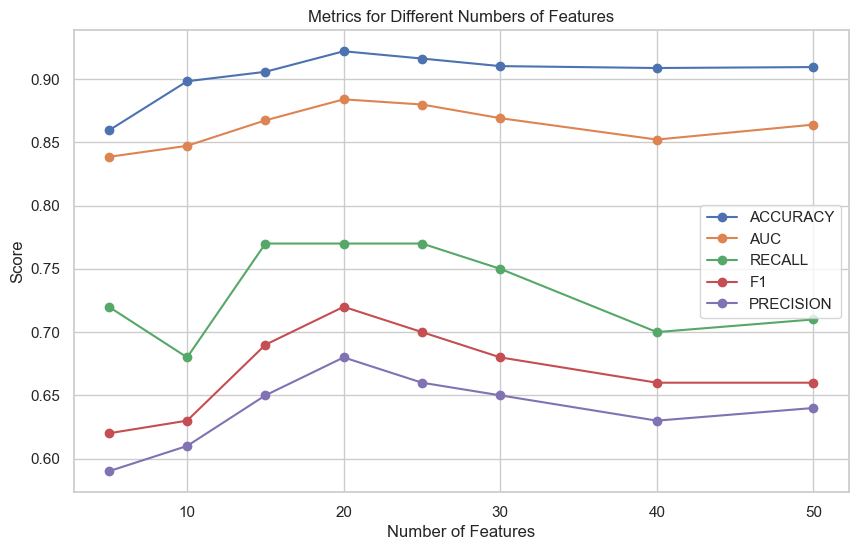}
  \caption{Metrics For Different Numbers Of Features}
  \label{fig:selected_num}
\end{figure}

The results indicate that the performance of the model varies depending on the number of features selected. Based on the metrics presented (e.g., AUC, Precision, Recall, F1, and Accuracy), it appears that the optimal number of features falls within the range of 15 to 25. Within this range, each of the metrics tends to peak, suggesting that this number of features balances complexity with predictive power. Selecting fewer features may reduce the model's predictive capacity, while too many features may introduce noise. the results are shown in Figure \ref{fig:selected_num}.

\subsection{From which category are the selected features chosen?}

\begin{figure}
  \includegraphics[width=\linewidth]{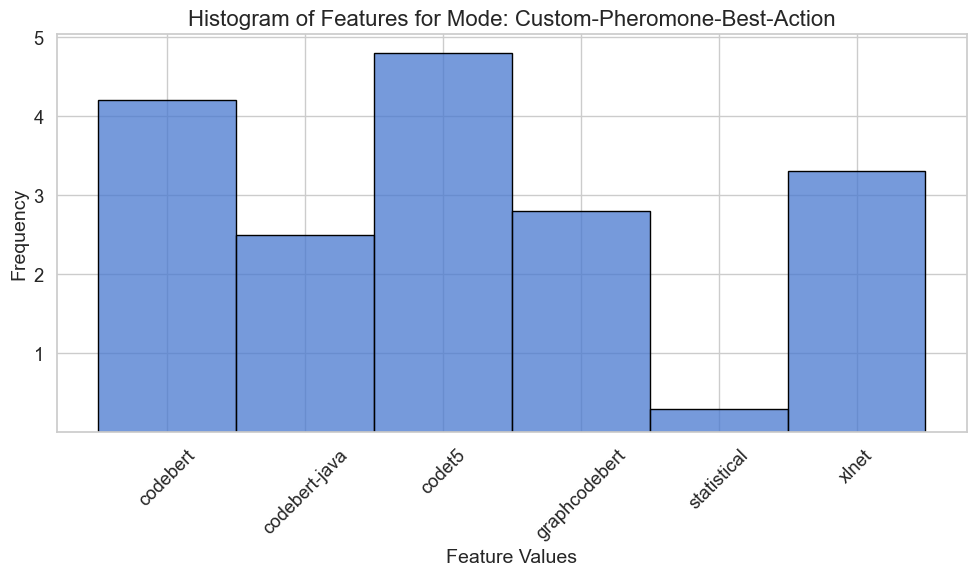}
  \caption{Number of Selected Features by the Best Performing Agent from Each Category}
  \label{fig:group}
\end{figure}

Figure \ref{fig:group} illustrates how the distribution of selected features highlights the significance of utilizing pre-trained models for feature extraction in SDP. The results obtained by the RL agent, when compared to baseline models, along with its preference for models like Codet5 and CodeBERT, underscore the importance of pre-trained models in the quality of the extracted features. Additionally, the average distribution of features chosen by the RL agent is well-balanced, indicating that each feature extraction unit complements the others effectively.

\subsection{How much does the Custom setting improve the approach compared to the Simple setting?}
\begin{figure}
  \includegraphics[width=\linewidth]{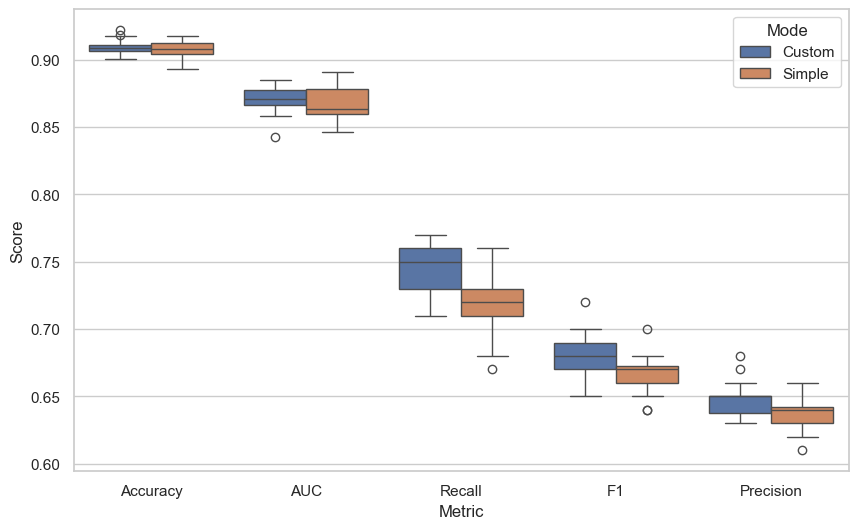}
  \caption{Comparison of Simple and Custom Modes in Different Metrics}
  \label{fig:simple-custom}
\end{figure}

We conducted experiments based on the two different settings of embedding usage: the Custom setting and the Simple setting. We examined the top-performing outcomes from experiments of equal size across varied pheromone table configurations. An independent samples t-test showed that the Custom setting significantly outperformed the Simple settings (t = 2.99, p = 0.0065). With a Cohen’s d of 0.61, this reveals a substantial and practically relevant benefit for the embedding configuration. An illustration of these experiments is shown in Figure \ref{fig:simple-custom}

\subsection{How much does the pheromone table improve the approach compared to the Vanilla form?}

\begin{figure*}[ht]
    \centering

    \subfloat[Simple setting]{%
        \includegraphics[width=0.48\linewidth]{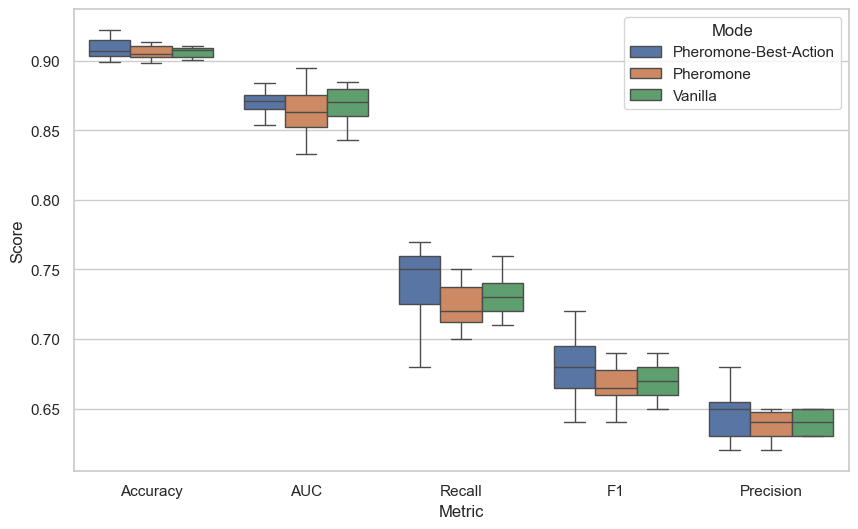}%
        \label{fig:baseline_simple}%
    }\hfill
    \subfloat[Custom setting]{%
        \includegraphics[width=0.48\linewidth]{settings_custom_comparison.png}%
        \label{fig:baseline_custom}%
    }

    \caption{Comparison of baseline settings.}
    \label{fig:baseline_comparison}
\end{figure*}

Based on the provided box plots \ref{fig:baseline_simple} and \ref{fig:baseline_custom}, the incorporation of a pheromone table significantly enhances SDP performance compared to the Vanilla approach. Results show that the Pheromone-Best-Action strategy achieves superior accuracy, AUC, and F1 scores over the Pheromone and Vanilla setups in the Custom setting. On the other hand, The Pheromone setup is dominant when using the Simple setting and outperforms Vanilla and Pheromone-Best-Action across all metrics. These findings highlight the potential benefits of combining RL and meta-heuristics in SDP.




\subsection{Answer to Research Questions}
\textit{RQ1: Can RL improve the performance of SDP models?}
The findings confirm that RL significantly enhances SDP performance. By utilizing the PPO algorithm, the proposed method dynamically selects the most impactful features from an extensive feature set. The use of RL reduces noise and ensures that only the most relevant features are passed to the classifier. Metrics such as AUC and F1-score showed marked improvement compared to baseline models, demonstrating the ability of RL to optimize predictive accuracy in diverse and complex datasets. These results align with recent studies, such as those exploring RL's ability to handle dynamic feature selection in defect prediction contexts. RQ1 is affirmative: RL proves to be an effective tool for enhancing defect prediction models.

\textit{RQ2: To what extent do pre-trained models enhance feature extraction to boost predictive performance in SDP?}

To answer RQ2, the quality of selected features underscores the effectiveness of using pre-trained models in feature extraction for SDP. Figure \ref{fig:selected_num} demonstrates that even with fewer selected features, their quality ensures robust predictive performance. This finding, along with the behavior of the agent demonstrated in Figure \ref{fig:group}, and its tendency to select from the features extracted by the pre-trained model, provides evidence for the efficacy of using these models in feature extraction. Consequently, given the remaining improvement gap, these results suggest exploring methods to enhance the quality of the extracted features.

\textit{RQ3: To what degree and in what manner can optimization of the agent's feature selection be improved?}
In this work, we introduced two approaches aimed at refining the optimization process of feature selection by RL: firstly, the implementation of an artificial embedding space, and secondly, the integration with a pheromone table inspired by ACO. The empirical results demonstrated the superior efficacy of both strategies when benchmarked against the Vanilla and Simple configurations, including scenarios where these were utilized in conjunction. The observed enhancements were substantial, with each approach contributing a meaningful increment in performance metrics. These findings offer valuable insights that could inspire the exploration of additional innovative approach combinations for further optimizing feature selection processes.

\subsection{Threats of Validity}
In this section, we discuss potential limitations and factors that might affect the validity of our results, including both internal and external threats.
\subsubsection{Internal Threats}
Internal threats to validity concern factors within the study that could influence the results. One potential internal threat is the selection of features, as the process may inadvertently favor specific types of features or overlook others that could contribute to model performance. Additionally, the performance metrics used (such as AUC, Precision, Recall, etc.) may vary with different dataset splits or sample sizes, potentially affecting the consistency of the results. The tuning of hyperparameters also introduces some variability, and different configurations could yield slightly different outcomes.
\subsubsection{External Threats}

External threats to validity address the generalizability of our findings to other datasets or contexts. Since this study is conducted on a specific dataset and within a controlled experimental setup, the results may not be fully transferable to other domains or datasets with different characteristics. Variations in feature distributions, sample sizes, or data collection methods in other datasets may impact the effectiveness of our feature selection methodology. Moreover, changes in the underlying data structure or the introduction of new features may necessitate re-evaluation of the optimal number and categories of features selected.








\section{Conclusion}

This paper introduces a novel framework for SDP that synergizes ensemble feature extraction and feature selection through RL. Central to our framework is the enhancement of the defect prediction process by defining a custom embedding space tailored to features and employing a pheromone table to guide the feature selection process. The proposed framework demonstrates robust adaptability to high-dimensional features and evolving software environments, as validated through extensive experiments on multiple projects within the PROMISE dataset. The results show consistent improvements in key performance metrics, including precision, recall, F1, and AUC, compared to traditional methods. This work paves the way for further research into cross-project defect prediction and the application of the framework in large-scale enterprise software systems, emphasizing its scalability and broader applicability.





\bibliographystyle{ieeetran}

\bibliography{cas-refs}


\end{document}